\documentclass[10pt]{article}
\usepackage[OE]{express}
\usepackage{siunitx}             
\usepackage{graphicx}            
\usepackage{xcolor}   
\usepackage{amsmath}
\usepackage[utf8]{inputenc}
\usepackage[T1]{fontenc}
\usepackage{booktabs}
\usepackage{float}
\usepackage{hyperref}  %
\hypersetup{colorlinks=true, citecolor=black,colorlinks, linkcolor=black, urlcolor=black}     
\usepackage{subfigure}
\usepackage{mathptmx} 

\hypersetup{pdfauthor={Matthew Aldous, Jonathan Woods, Andrei Dragomir, Ritayan Roy and Matt Himsworth}, pdftitle={Carrier frequency modulation of an acousto-optic modulator for laser stabilization}, pdfkeywords={laser spectroscopy, acousto-optic modulator, rubidium spectroscopy }}
\DeclareSIQualifier\peak{p}      
\DeclareSIQualifier\peakpeak{pp} 
\DeclareSIUnit\sample{S}
\DeclareSIUnit\gauss{G}          

\let\oldhat\hat	                                          
\renewcommand{\hat}[1]{\boldsymbol{\oldhat{\mathbf{#1}}}} 

\let\footnote=\endnote

\begin{document}
\title{Carrier frequency modulation of an acousto-optic modulator for laser stabilization}
\author{Matthew Aldous, Jonathan Woods, Andrei Dragomir, Ritayan Roy and Matt Himsworth\authormark{1}}
\address{School of Physics and Astronomy, University of Southampton, Highfield, Southampton, SO17 1BJ, United Kingdom}
\email{\authormark{1}m.d.himsworth@soton.ac.uk}

\vspace{10pt}

\date{\today}

\begin{abstract}
The stabilization of lasers to absolute frequency references is a fundamental requirement in several areas of atomic, molecular and optical physics. A range of techniques are available to produce a suitable reference onto which one can `lock' the laser, many of which depend on the specific internal structure of the reference or are sensitive to laser intensity noise. We present a novel method using the frequency modulation of an acousto-optic modulator's carrier (drive) signal to generate two spatially separated beams, with a frequency difference of only a few MHz. These beams are used to probe a narrow absorption feature and the difference in their detected signals leads to a dispersion-like feature suitable for wavelength stabilization of a diode laser. This simple and versatile method only requires a narrow absorption line and is therefore suitable for both atomic and cavity based stabilization schemes. To demonstrate the suitability of this method we lock an external cavity diode laser near the $^{85}\mathrm{Rb}\,5S_{1/2}\rightarrow5P_{3/2}, F=3\rightarrow F^{\prime}=4$ using sub-Doppler pump probe spectroscopy and also demonstrate excellent agreement between the measured signal and a theoretical model. 
\end{abstract}


\bibliographystyle{osajnl_w_doi}
\bibliography{AOMref}

\section{Introduction}\label{intro}

Frequency stabilization of a laser to a known reference is a common requirement in a number of applications, such as atomic laser cooling, absorptive sensing and precision spectroscopy. The most stable frequency references are atomic transitions and numerous techniques exist to obtain a suitable `error signal' which can be electronically fed back to the laser to correct for frequency drift. Effective methods include frequency modulation spectroscopy (FMS) \cite{bjorklund1979}, dichroic atomic vapour laser lock (DAVLL) \cite{Corwin1998}, polarization spectroscopy (PS) \cite{Wieman1976} and modulation transfer spectroscopy (MTS) \cite{McCarron2008}, all of which can be used with sub-Doppler pump-probe methods to obtain very effective stabilization signals. In many cases one would prefer to avoid modulated sidebands on the laser spectrum and so there is interest in modulation-free spectroscopy, or techniques in which the modulation is confined to the spectroscopy system. The available techniques can be separated into two distinct methods: \emph{phase-detection} spectroscopy (as used in FMS and MTS) which detects variation in the phase relationship between modulation sidebands on different sides of an absorption feature, and \emph{frequency differential} spectroscopy (found in DAVLL and PS), where two absorption spectra separated (spatially, temporally, or both) with a frequency shift are subtracted to generate the error signal.
	
While both DAVLL and PS are very effective, they produce the differential frequency shift using the internal structure of the atoms under investigation, and this may not always be practical to access in a species where the electronic structure is not suitable or if one wishes to stabilize to a non-atomic reference. DAVLL achieves this by the Zeeman effect, and polarization spectroscopy through optical pumping; both measure the offset spectra via orthogonal polarization states, and balanced detection of the differential signals provides common-mode rejection, greatly reducing the effect of laser intensity noise on the spectra. An alternative spectroscopic method, presented here, uses the balanced detection between two spatially- and frequency-separated laser beams to provide a dispersion-shaped signal across an absorption feature. The beams are produced via the carrier modulation of an acousto-optic modulator's (AOM) drive frequency and we demonstrate an example application of the method using sub-Doppler pump-probe spectroscopy of rubidium (Rb) vapour as a wavelength reference, but this versatile technique can be applied to any spectroscopic feature of appropriate width.

\section{Carrier frequency modulation of an acousto-optic modulator}\label{theory}
In laser cooling experiments, it is common to use AOMs to switch trapping and optical-pumping beams on and off within short timescales, and to provide a tunable frequency offset. An AOM introduces an angular deviation of the optical path, capable of providing several deflected beams which are frequency shifted from the zeroth-order by harmonics of the AOM's carrier frequency. Several schemes \cite{Zhang2009,VanOoijen2004} have been proposed to use AOMs to produce dispersion-shaped spectroscopic features, all using the differential method between the various diffracted orders or involving multiple AOMs. It is understood that to obtain a well-resolved locking signal, the frequency difference between absorption spectra should be less than the width of the feature of interest. Additionally, the smaller the frequency difference, the steeper the error signal gradient and therefore the better-resolved the reference.
	
Commercial AOMs can be found with operating frequencies of several tens to hundreds of MHz, and for many applications this combination of parameters leads to a large capture range and good stability. The most demanding stabilization, however, requires locking to a very narrow absorption feature with a linewidth of a few MHz or below and so the frequency offset between diffracted orders in the AOM system (equal to $n\times f_0$ where $n$ is the order index and $f_0$ is the AOM carrier frequency) is usually far too large. Typically AOMs are operated with a single drive frequency, but if this carrier is modulated, it provides sidebands on the diffracted beam with a bandwidth of a few tens of MHz. This method can be an economic alternative to Electro-Optical Modulation (EOM) to produce frequency sidebands and has been used as to replace EOMs in Modulation Transfer Spectroscopy \cite{Negnevitsky2013}. This `carrier modulation' is a simple method to generate spatially- and frequency-separated spectroscopic probe beams, which may be detected in a balanced manner and subtracted electronically to obtain the necessary error signal. For our application, we see the common mode rejection and simplicity of the optical and electronic set-up as attractive properties of this apparatus.

\begin{figure}[t]
\centering
\includegraphics[width=0.8\textwidth]{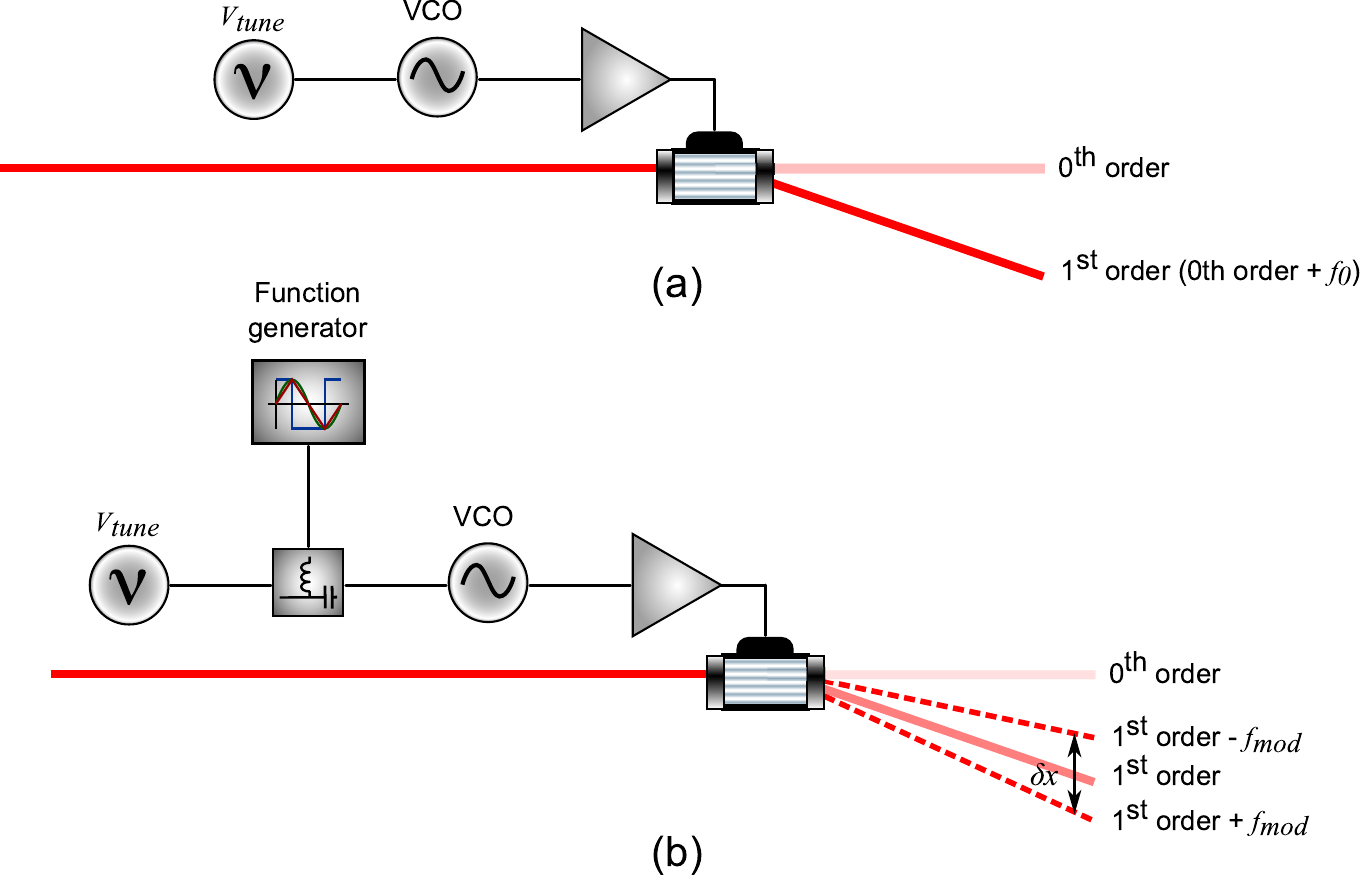}
\caption{Diagram of diffraction through the AOM in a) normal operation with a drive frequency $f_0$ and b) with the drive signal modulated at $f_{mod}$.}
\label{diffraction-fig}
\end{figure}

The angle of divergence between the $0$\,th and $1$\,st diffracted order from an AOM driven with a carrier frequency $f_0$ is given by \cite{Donley2005}:
	
\begin{center}
$\theta =\frac{f_0\lambda}{v_g}$,
\end{center}
		
where $\lambda$ is the wavelength of the incident beam and $v_g$ is the acoustic velocity in the AOM crystal. If the AOM is driven by two frequencies with a difference of 5\,MHz, the angular separation of the beams would be $\simeq$0.05$^\circ$ (for $\lambda=780\,$nm and using a TeO$_2$ crystal with $v_g=4200$\,m/s along the (110) plane \cite{Ohmachi1972}). While these diverging beams may be picked off and directed onto individual balanced detectors, a segmented photodiode may be more convenient when monitoring beams with small separations. For a photodiode with segments separated by a $200\,\mu$m gap, the optical path length from AOM to detector must be at least 230\,mm in order for more than half of each beam spot to fall on the correct segment. This length scale is generally acceptable for most experiments. The distinction between the carrier (drive) frequency and the sidebands produced by modulation of the $V_{\mathrm{tune}}$ signal is presented in Figure \ref{diffraction-fig}.
		
The signal strength for small separations would be equal to the difference in overlap of the beam widths, $\delta x$, on the detector. Assuming a Gaussian beam shape the signal as a function of frequency shift $\Delta$ would be proportional to:
\begin{center}
$I_{min}(\Delta)=I_0\left(1-\exp\left[-\left(\frac{f_0\lambda\Delta}{2 v_g \delta x}\right)^2\right]\right)$,
\end{center}
where, $I_0$ is the maximum optical intensity. The upper bound in frequency would be defined by the AOM analog modulation bandwidth, the size of the detector or both. The former is a fundamental boundary and is dependent on the $1/e^2$ beam width within the AOM, $d$:
\begin{center}
$I_{max}(\Delta)=I_0\exp\left[-\frac{1}{8}\left(\frac{\pi d \Delta}{v_g}\right)^2\right]$.
\end{center}

\section{Experiment}\label{expt}

The laser source used is a $780\,$nm external cavity diode laser (ECDL) built based on established designs \cite{Arnold1998, Hawthorn2001}, which we use for laser cooling of atomic rubidium, and therefore must be locked precisely to a given electronic transition (specifically $^{85}$Rb D$_2 (5S_{1/2}\rightarrow5P_{3/2})$) to better than $1\,$MHz. Figure \ref{diffraction-fig} shows the layout of the spectroscopy system. An acousto-optic modulator (\emph{Gooch \& Housego} FS310-2F-SU4), with a center carrier frequency $f_0 = 310$\,MHz, is aligned in the Bragg regime such that only a single diffraction order is produced. Note that the choice of AOM carrier frequency is arbitrary here, and that similar results were also produced using an AOM operating at 80\,MHz, the only practical difference being the corresponding AOM modulation bandwidths.

\begin{figure}[b]
\centering
\includegraphics[width=0.8\textwidth]{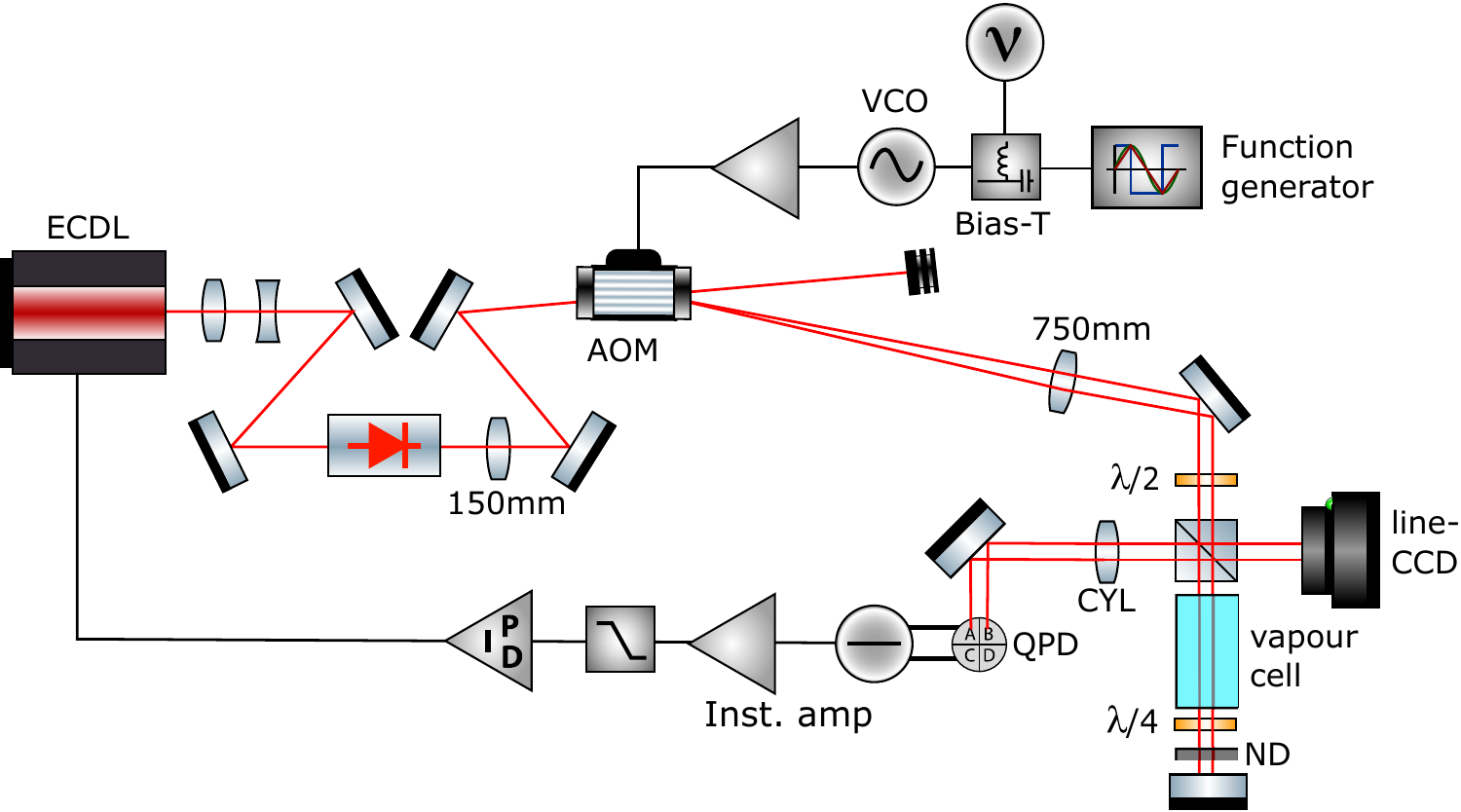}
\caption{The spectroscopy apparatus. The beam from an external cavity diode laser is collimated, isolated and focused through a 310\,MHz acousto-optic modulator (AOM). This is driven by an amplified voltage controlled oscillator (VCO). The VCO is tuned with a square wave modulation combined with a DC offset using a bias-tee. The multiple diffracted beams from the AOM are passed through a retro-reflected Doppler-free spectroscopy system including a neutral-density (ND) filter and half- and quarter-waveplates ($\lambda/2$, $\lambda/4$). The retro-reflected signal can either be obtained using polarizing optics, or via back-reflection through the AOM and simplifying the apparatus. The beam shape is also monitored using a line scan CCD (\emph{Thorlabs} LC1-USB). The final detection is achieved on a quadrant photodiode (QPD) after concentration of the sideband beams by a cylindrical lens (CYL).}
\label{setup} 
\end{figure}

One may produce the necessary RF fields either by modulating the carrier or by using two distinct carrier frequencies (at $f_0\pm \Delta$). Both methods produce similar results and all of the following data was obtained using the former method by simply modulating the tuning port of a voltage controlled oscillator (VCO) to produce the RF sidebands. We apply a square-wave RF waveform to the VCO (\emph{Mini-Circuits} ZX95-330-S+) tuning port using a bias tee (\emph{Mini-Circuits} ZX86-12G+). The VCO output is then amplified to 27$\,$dBm (\emph{Mini-Circuits} TVA-11-422) and fed to the AOM.  By modulating the tuning pin of the VCO,  $\Delta$ is defined by the modulation \emph{amplitude}, not the frequency $f_{mod}$, with a dependence of $\sim9$\,MHz/V for a square wave signal. The modulation frequency $f_{mod}$ need only lie within the bandwidth of tuning port and results in a modulation `noise' in the detected signal which must be filtered out. The use of two distinct drive frequencies avoids this noise but requires a dedicated waveform generator to maintain the frequency difference between the components. The resulting pair of beams is then collimated and passed through a sub-Doppler pump-probe apparatus as shown in Figure \ref{setup} and also directed onto a line-scan charge-coupled device CCD (\emph{Thorlabs} LC1-USB). For simplicity we use the retro-reflected pump-probe configuration with the probe beam picked off with polarization optics.

The two beams are detected using a quadrature photodiode (QPD, \emph{Centronic} QD7-5T), with two quadrants on each side summed together individually, before subtraction of the pairs to generated the differential signal. We focus the beams onto the detector using a cylindrical lens, oriented to increase optical capture on the sensing region without the reduction of spot separation associated with lateral focusing. The QPD sections are individually biased and the output of two horizontal sections are subtracted using an instrumentation operational amplifier with 20\,dB gain, before passing through a 100\,kHz low-pass filter, mandatory in order to eliminate systematic noise introduced by the VCO tuning pin modulation frequency. As the modulation frequency has little effect on the spectra one can choose a combination of filter and modulation frequency to suit the stabilization circuit bandwidth.
			
\section{Results}

\begin{figure}[b]
\centering
\includegraphics[width=0.6\textwidth]{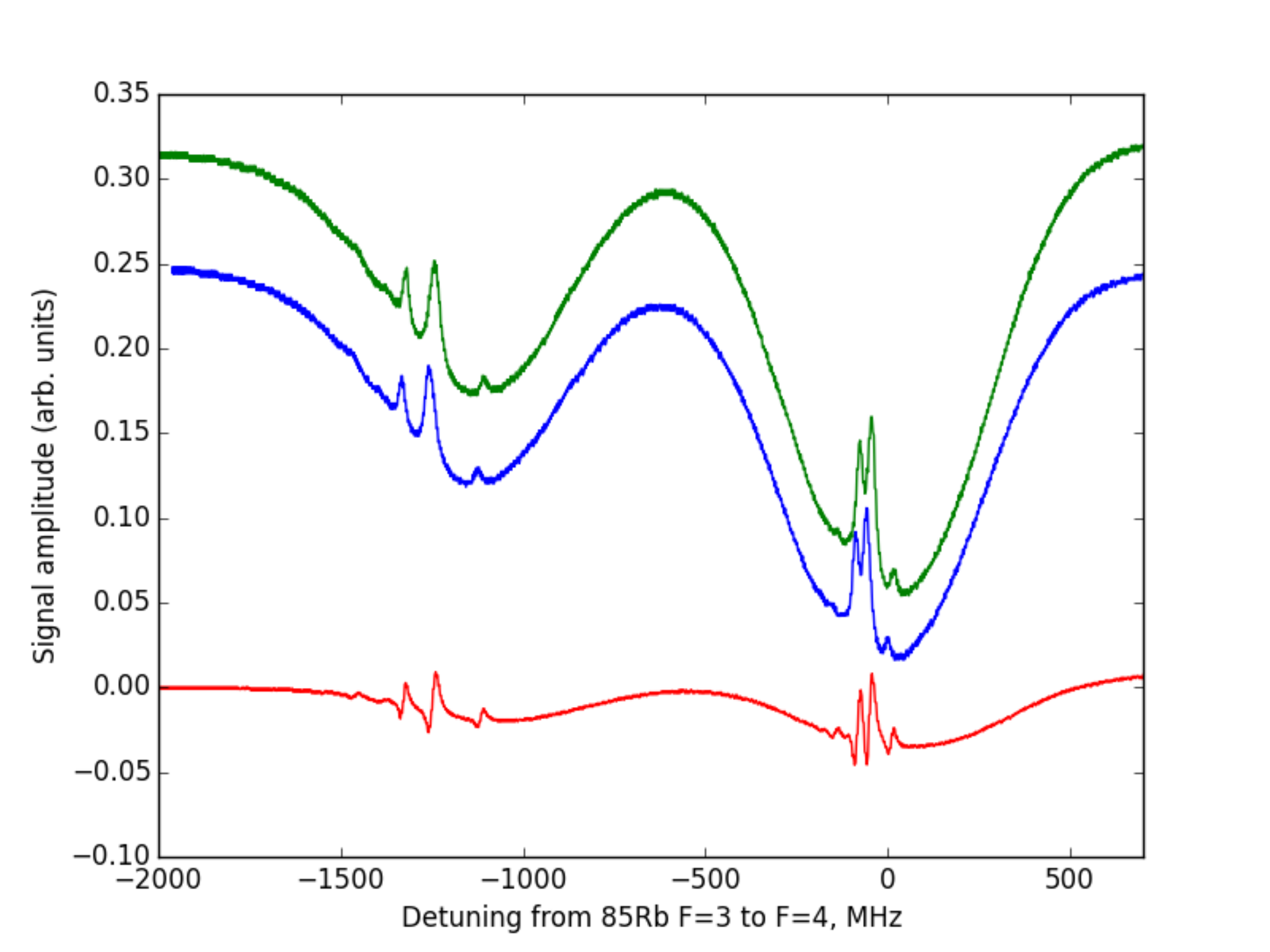}
\caption{Absorption spectra (blue and green curves) from each half of the QPD and (red curve) the associated error signal derived from them.}
\label{spectra} 
\end{figure}

An example of sub-Doppler spectra detected by each half of the QPD are shown in Figure \ref{spectra}, recorded by zeroing the input of each into the instrumental amplifier in turn. The horizontal axis has been scaled using the known frequencies of each absorption peak fitted using a 4th order polynomial to mitigate non-linearity of the piezoelectric tuning. We also plot the direct subtraction of the absorption spectra without using the instrumentation amplifier, where the subtracted signal is very similar to an FMS or PS spectrum.  The signal to noise ratio of the subtracted data is much higher because any intensity noise in the laser is common-mode to both beams and thus subtracted with the high bandwidth (2\,MHz) instrumentation op-amp. We find that the subtracted signal is remarkably insensitive to variations in the laser power, other than changing the overall signal strength around zero.

\begin{figure}[b]
\centering
\includegraphics[width=0.6\textwidth]{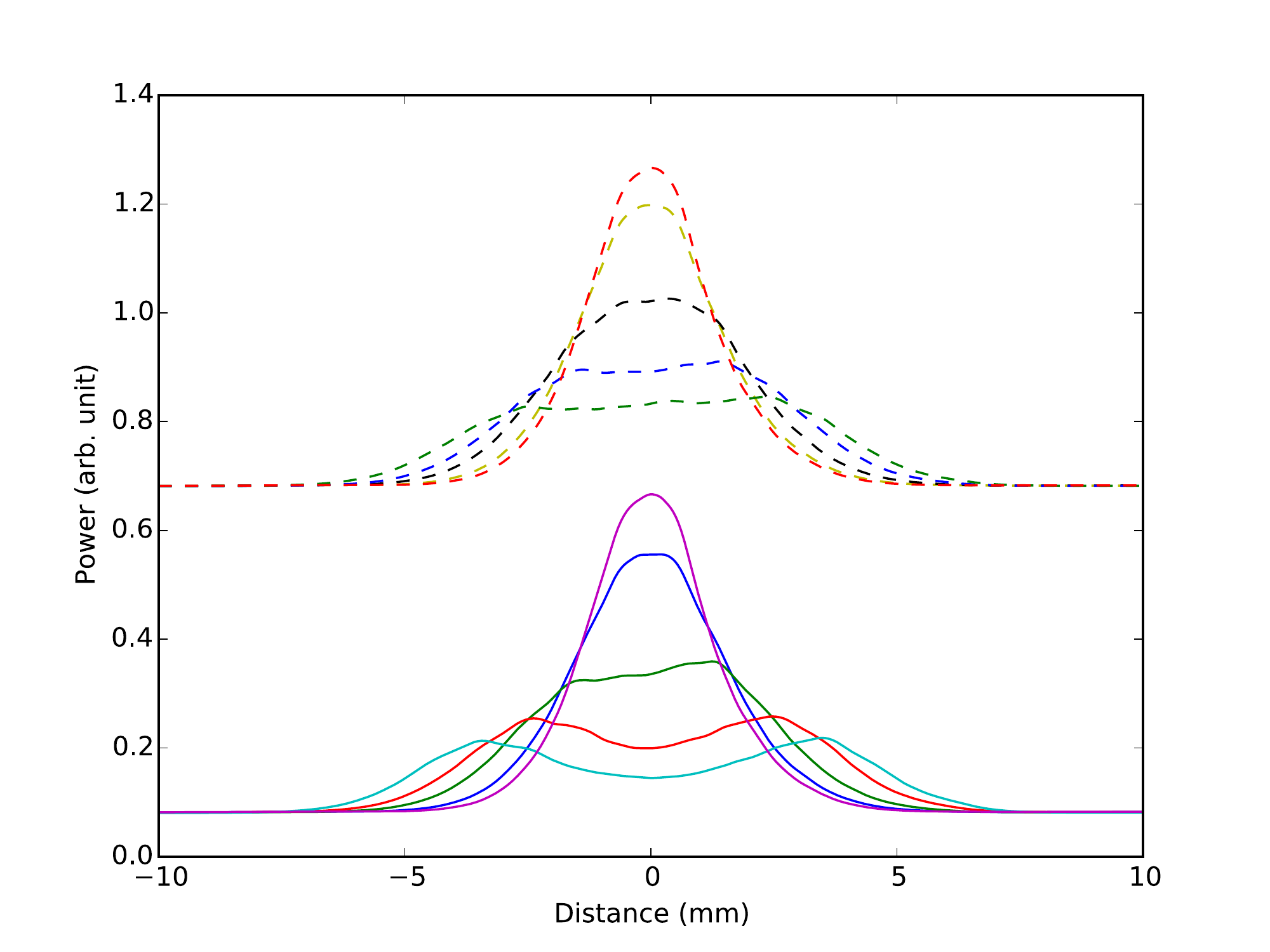}
\caption{The beam shape of the first order diffracted beam with different modulation modes, measured using a line-scan CCD. The upper dashed trace is produced using a sine-wave modulation of the VCO tuning voltage, the lower solid traces used a square-wave modulation. The square-wave spectra show more power in the sidebands compared to the sinusoidal modulation because the tuning voltage does not have a significant carrier component (310\,MHz at 0\,mm in this demonstration).}
\label{beams} 
\end{figure}

	

\begin{figure}[tb]
\centering
\subfigure[]{\includegraphics[width=0.49\textwidth]{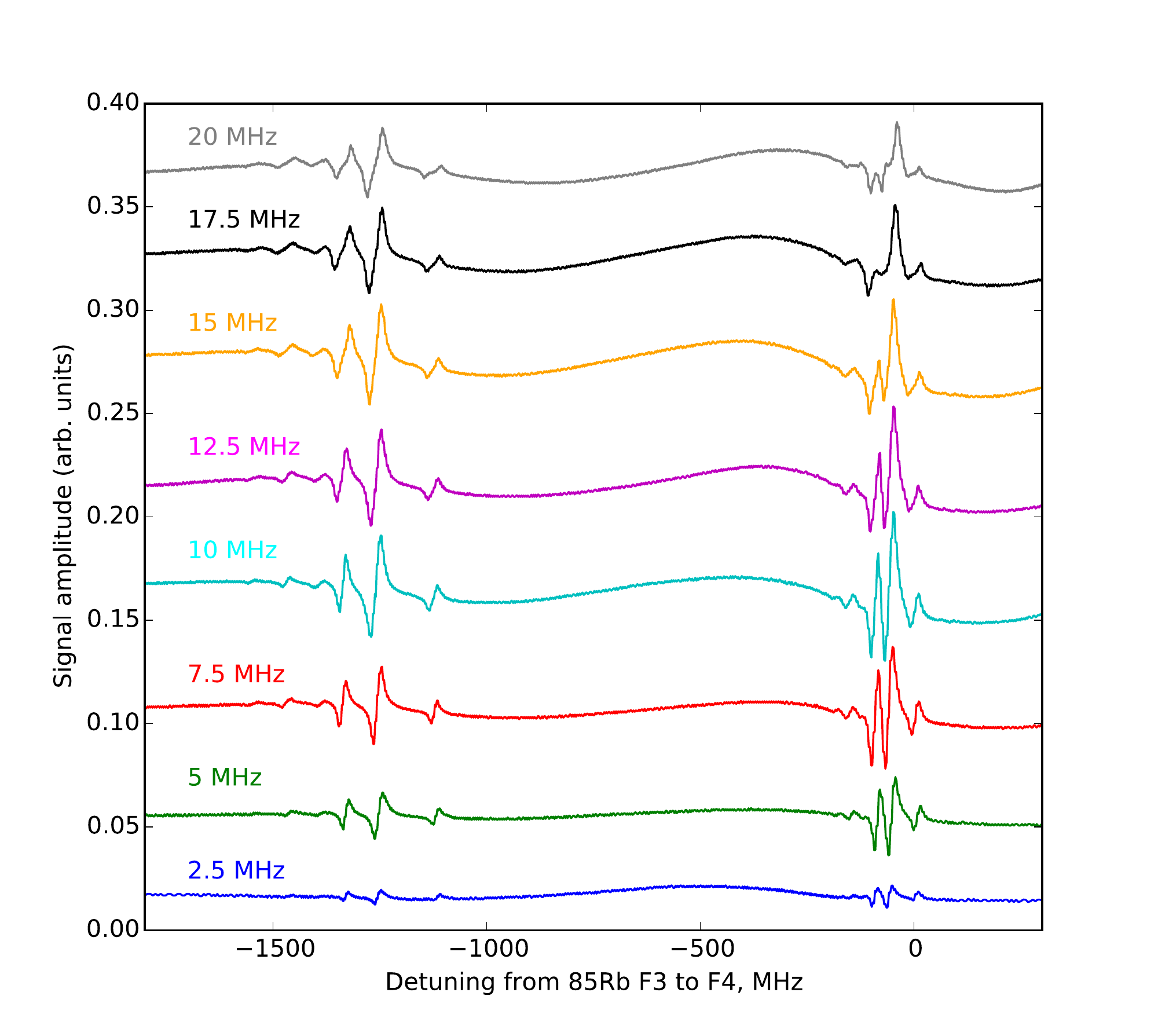}\label{voltdata}}
\subfigure[]{\includegraphics[width=0.49\textwidth]{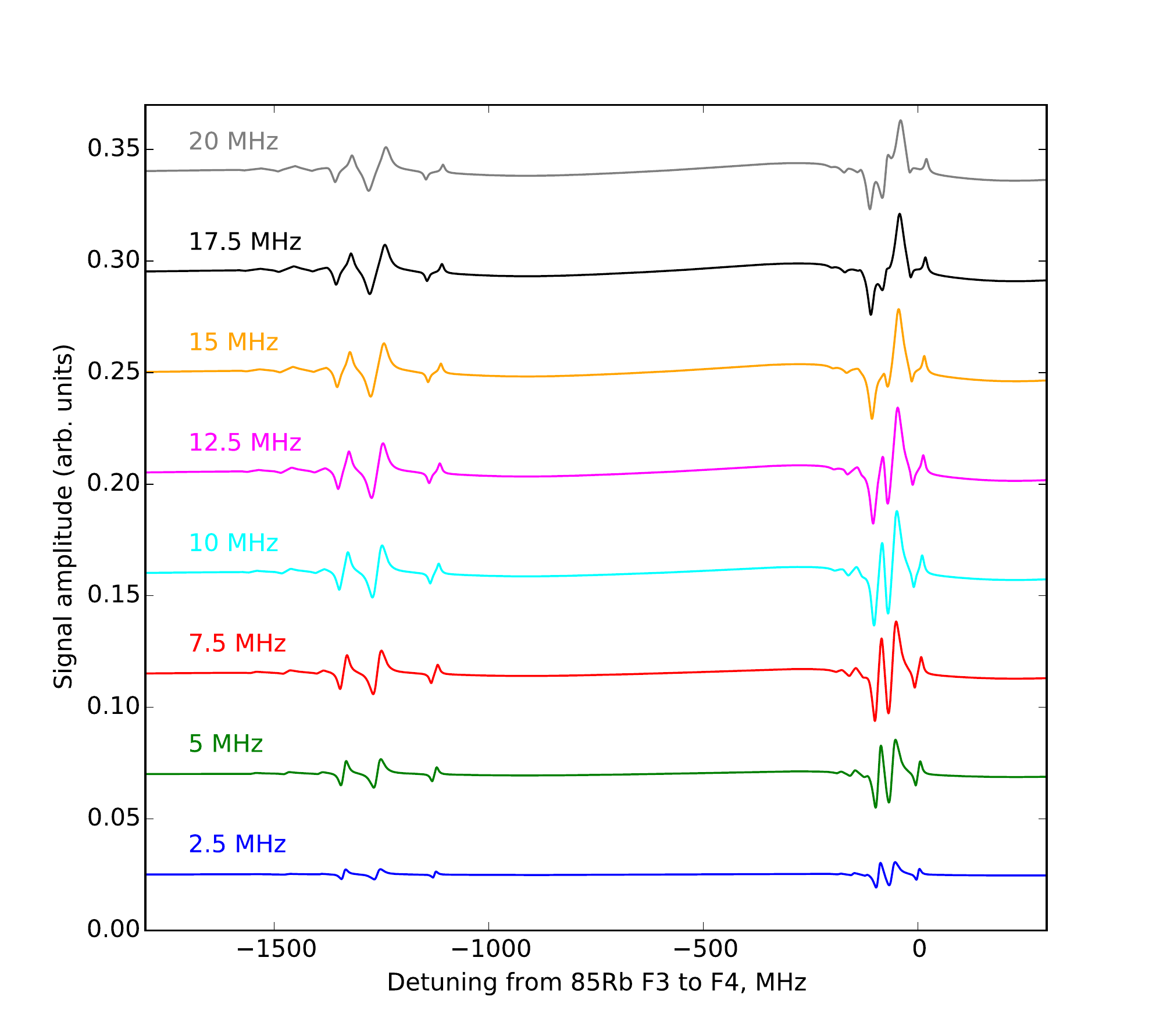}\label{voltmodel}}
\caption{Measured error signal at the detector for different modulation frequencies (a) compared to a numerical model with no free parameters\cite{Himsworth2010} (b). In our system the optimum regime in both cases is in the vicinity of 10\,MHz.}
\label{volt}
\end{figure}
	
\begin{figure}[b]
\centering
\subfigure[]{\includegraphics[width=0.49\textwidth]{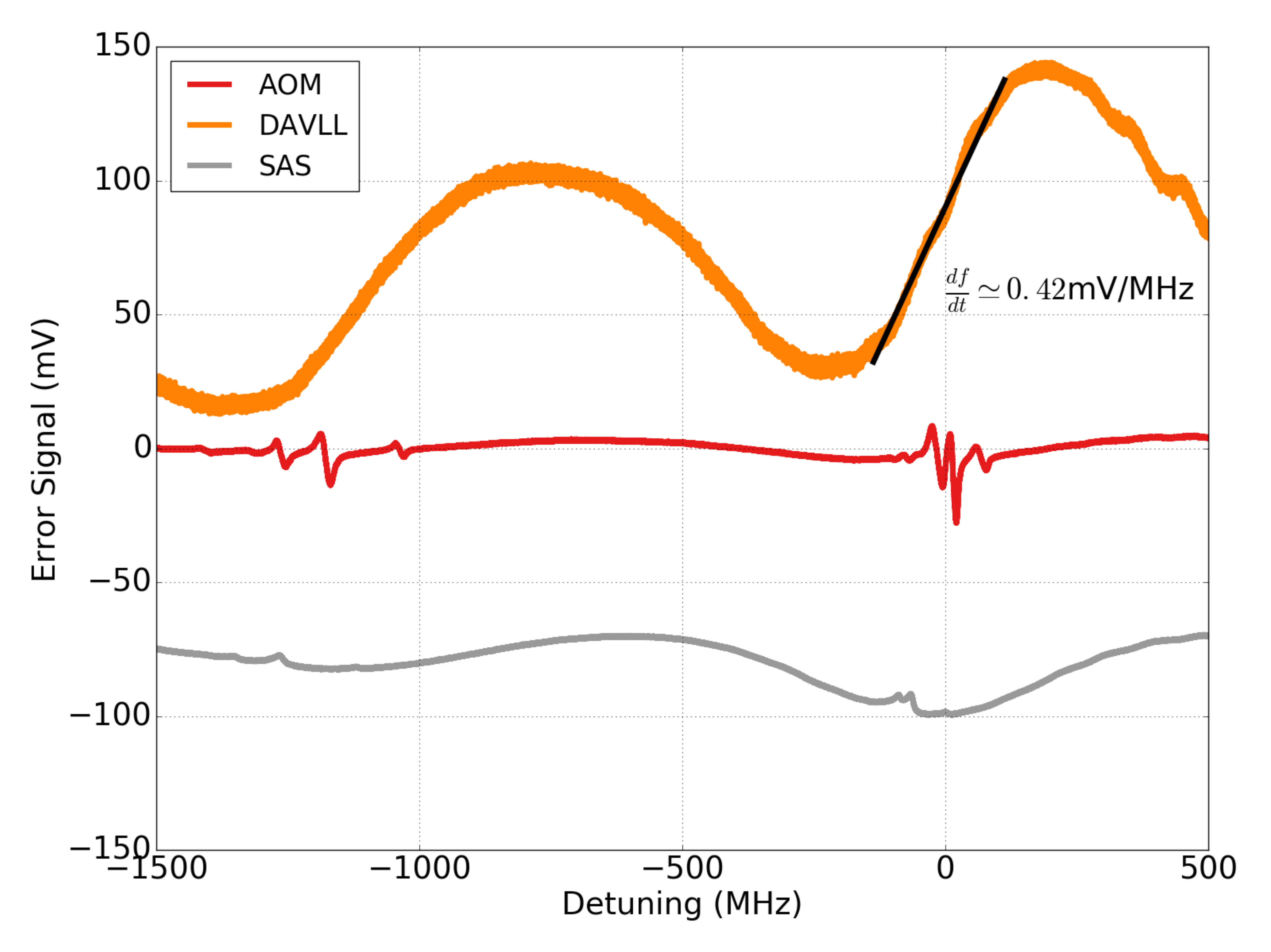}\label{fig:time-calibration}}
\subfigure[]{\includegraphics[width=0.49\textwidth]{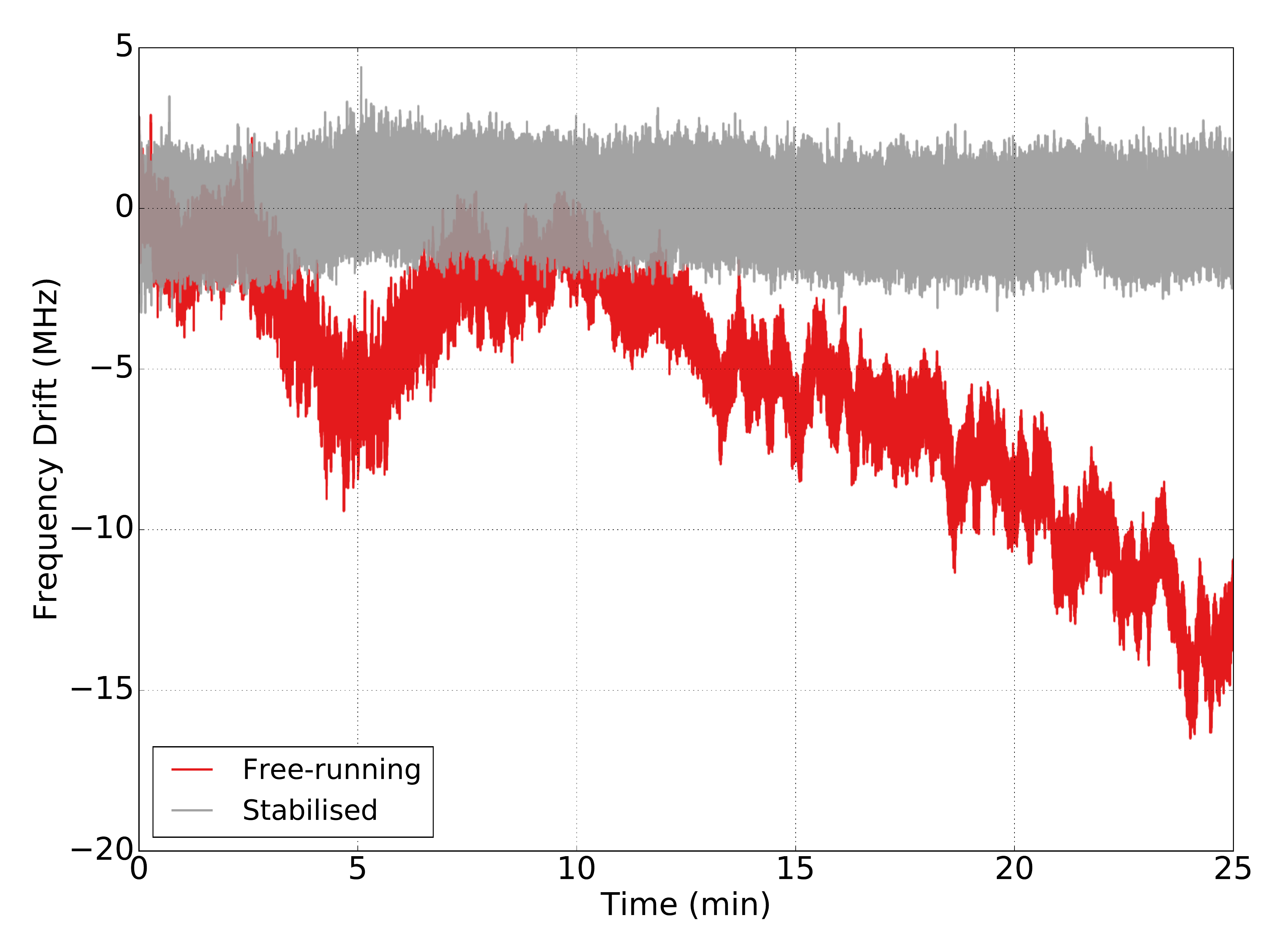}\label{fig:locking-comparison}}
\caption{a) Absorption spectrum and error signals used in the time-domain analysis. The vertical offsets of the signals are for clarity only, and the black line represents a pseudo-linear fit to the locking region, with an approximate gradient of 0.42\,mV/MHz. b) Data showing performance of the laser system across two separate 25-minute runs, with the AOM system in free-running (FR) and stabilized (L) mode. The final drifts are $\Delta f_{FR}=54\pm3$\,MHz and $\Delta f_{L}=0.66\pm2$\,MHz (r.m.s.).}
\end{figure}

We have explored varying both the dither frequency, amplitude, and waveform shape. We find very little variation between sine and square waves, except at high frequencies where sinusoidal modulation causes less distortion in the final signal due to the tuning pin bandwidth filtering the higher modes of the square wave. Modulating with a sinusoidal signal produced beam profiles with an inferior resolution, as well as causing the sideband frequency separation to be proportional to the r.m.s. amplitude, as shown in Figure \ref{beams}. Within the frequency range from 200\,kHz to 5\,MHz we see negligible variation in the spectra for both modulation waveforms. 

Figure \ref{voltdata} shows the variation with sideband separation (proportional to modulation amplitude), from 500kHz to 20MHz together with a prediction (shown in Figure \ref{voltmodel}) using a theoretical model with no free parameters \cite{Himsworth2010}. The optimum lineshape, where the error signal is most linear across resonance is found around 8-12\,MHz sideband separation. This is the linewidth of the sub-Doppler absorption features used (which is slightly broader than the natural linewidth), as expected from the theoretical model. At lower separations the smaller differences between spectra significantly weaken the derived signal, and at higher frequencies the separations are greater than the sub-Doppler linewidths so the different absorption and cross-over peaks begin to overlap.

To test the suitability of this technique for laser stabilization, a parallel DAVLL setup using the same laser passing through a different vapor cell was used to characterize the long-term drift of a laser locked using this technique \cite{Aldous2016}. The DAVLL signal, which in our case is only sensitive to the Doppler-broadened spectral features, was frequency shifted by a further AOM such that the zero in its error signal was situated very close to the center of the reference transition (a crossover resonance in the vicinity of $^{85}\mathrm{Rb}\,5S_{1/2}\rightarrow5P_{3/2}, F=3\rightarrow F^{\prime}=4$). This provided a diagnostic signal proportional to any drift, even if the system was far off-resonance. The error signal used was supplied to a proportional-integral-derivative (PID) controller which in turn fed back to the laser diode current and the piezo-mounted external grating. An overview of the spectra during a single laser sweep is shown in Figure \ref{fig:time-calibration}, which includes a SAS spectrum alongside the modulated AOM and DAVLL error signals.

The drift of the ECDL system was measured over 25 minutes in both free-running and locked modes of operation, as shown in Figure \ref{fig:locking-comparison}. The maximum drift in DAVLL signal indicates the free-running laser naturally drifts on the order of $13\pm3$\,MHz during the 25\,min recorded period ($\simeq50$\,MHz per hour) which is comparable to the stability of similar lasers tested in the literature \cite{Matsubara2005}. Once the laser is locked there is no significant drift with a r.m.s. frequency variation of 0.66\,MHz, which is approximately equal to the laser linewidth \cite{himsworth2009coherent}. The lock remains remarkably secure, even within a noisy laboratory environment and with vibration of the optical breadboard.

\subsection{Application to frequency modulation spectroscopy}
\begin{figure}[t]
\centering
\includegraphics[width=0.6\textwidth]{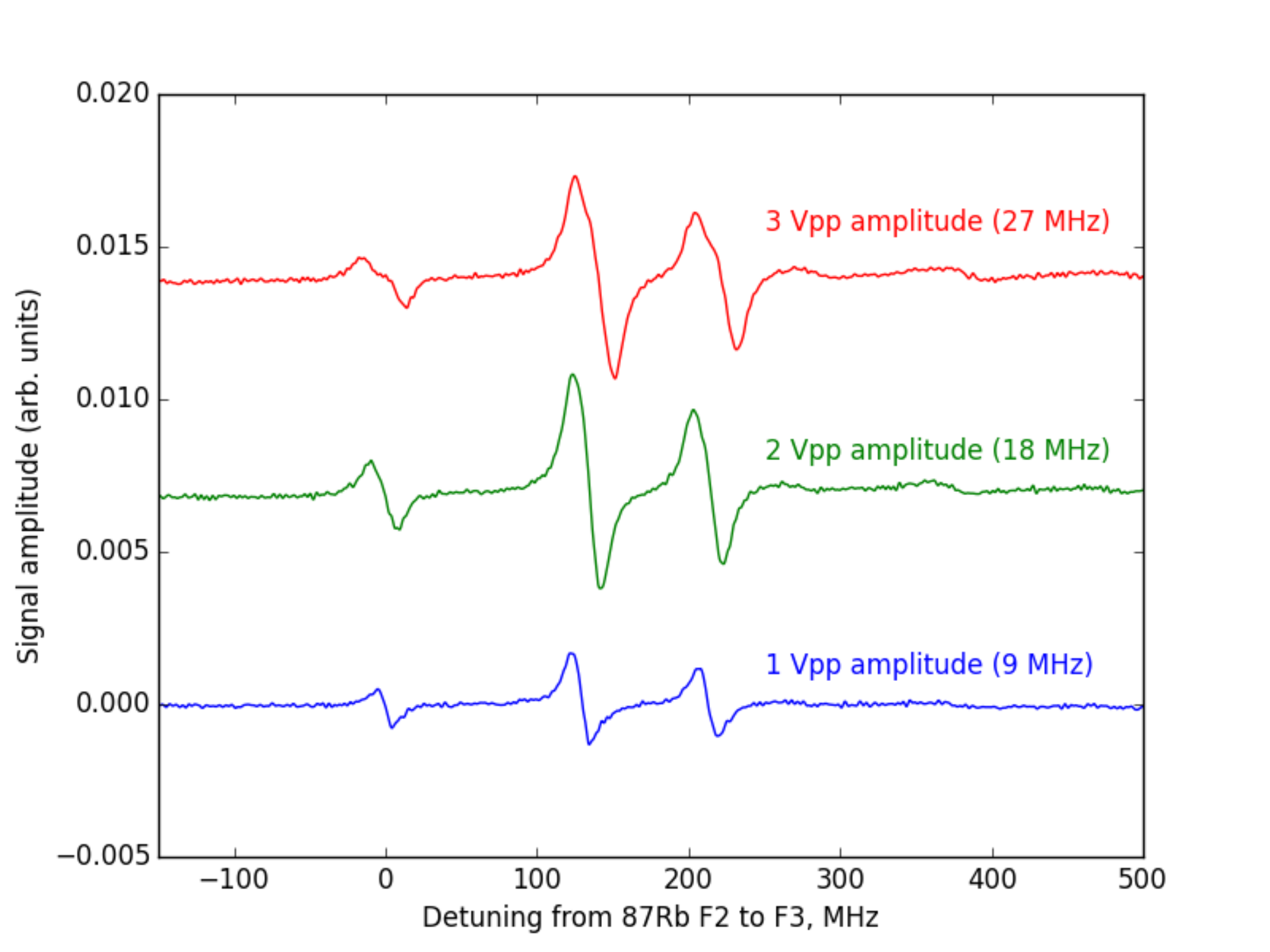}
\caption{Variation of the demodulated error signal with the VCO modulated with a square wave at 3\,MHz as the sideband separation is swept from 9 to 27\,MHz.}
\label{f-mod-data}
\end{figure}

Although we have focused on differential methods to obtain the error signal, the use of a modulated tuning voltage of the VCO offers an interesting version of frequency modulation spectroscopy. In generating the two 1st-order beams we modulate the VCO with a waveform whose amplitude defines their frequency separation, and the frequency of the modulation is essentially a noise source which is filtered out. However, if a single detector is used and we demodulate at the same frequency with which the VCO tuning port is driven, then we find that it is possible to produce a FMS error signal whose sideband frequency is decoupled from the demodulation frequency. Figure \ref{f-mod-data} shows a selection of spectra with a constant modulation frequency but a variable sideband separation (via the tuning port modulation \emph{amplitude}). The signal strength of FMS spectra typically reduces at higher modulation frequency with narrow absorption features \cite{silver1992frequency}, however one requires the modulation frequency to be above the noise bandwidth of the laser (typically below 1 or 2\,MHz for a external cavity diode laser). Therefore it may be of interest to exploit this element of the technique if a specific sideband separation, independent of the demodulation frequency, is required.

\section{Discussion} 

We find the modulation frequency of the VCO tuning port to have little effect on the spectra in the range 200kHz to 5MHz: an operating range determined by the overlap in the bandwidths of the bias tee and the VCO tuning. The use of square or sinusoidal modulation waveshapes also has little effect on the spectra other than changing the sideband separation, however the the use of square waves allows one to alter the duty cycle of the modulation and thus produce small frequency offsets from the absorption reference. 

One weakness of the technique proposed here is its sensitivity to fluctuations in the pointing direction of the beam emerging from the AOM caused by pressure fluctuations in the laboratory, since the ratio of optical intensities falling on each detector segment may vary, thus producing a change in the DC offset. Beyond shielding the apparatus, the presence of the focusing lens in front of the detector, but at a distance less than the focal length, serves to mitigate this by reducing the beam spot size on the detector in comparison to the sensor area.

Since the VCO in our apparatus is not stabilized to the AOM's center frequency, any slow drift results in a variation of power in each beam and thus a drift in the error signal's DC offset. Therefore, for long term stabilization a precision voltage reference is necessary, or a second QPD can be used to monitor the power in each beam, feeding back to stabilize the VCO (in much the same manner as the spectroscopic signal is used to stabilize the laser).

The apparatus can be made more compact if one discards the beam-splitting cube and allows the retro-reflected probe beam to pass back through the AOM after which its undeflected component may be focused on the QPD.  

\section{Conclusion}

A new method for wavelength stabilization of a laser diode has been demonstrated that depends on the carrier modulation of the AOM drive frequency to provide spatially and spectrally separated sidebands. These are used to jointly probe an absorption feature and the difference in the detected signals produced an error signal suitable for locking. A simple RF electronic system was also presented to produce the correct RF drive signal via the modulation of a VCO tuning port.  An advantage of this method is its insensitivity to laser intensity noise, background electrical or magnetic fields, and optical polarization. Therefore it is suitable for both atomic and cavity wavelength references, especially where narrow absorption features are required for the highest precision. The technique was used to lock an external cavity diode laser to a sub-Doppler absorption line in rubidium and the measured stability, at one part in $10^9$, is suitable for cold atoms experiments.

\section*{Funding}
This work was supported by funding from RAEng, EPSRC, and the UK Quantum Technology Hub for Sensors and Metrology under grant EP/M013294/1.

\section*{Acknowledgements}
We would like to thank Paul Martin, Sanja Barkovic for their help in building up the apparatus, and Tim Freegarde for useful discussions and for the loan of certain pieces of equipment. 

\end{document}